    \documentclass[10pt]{iopart}
    
    \usepackage{iopams} 
    \usepackage{hyperref} 
    \usepackage{graphicx}

\begin{document}

    \title[Solid Angle and Knotted Fields]{Maxwell's Theory of Solid Angle and the Construction of Knotted Fields}

    \author{Jack Binysh$^1$ and Gareth P Alexander$^2$}
    \address{$^1$ Mathematics Institute, Zeeman Building, University of Warwick, Coventry, CV4~7AL, UK.}
    \address{$^2$ Department of Physics and Centre for Complexity Science, University of Warwick, Coventry, CV4~7AL, UK.}
    \eads{\mailto{j.binysh@warwick.ac.uk}, \mailto{g.p.alexander@warwick.ac.uk}}

    \begin{abstract}
    We provide a systematic description of the solid angle function as a means of constructing a knotted field for any curve or link in $\mathbb{R}^3$. This is a purely geometric construction in which all of the properties of the entire knotted field derive from the geometry of the curve, and from projective and spherical geometry. We emphasise a fundamental homotopy formula as unifying different formulae for computing the solid angle. The solid angle induces a natural framing of the curve, which we show is related to its writhe and use to characterise the local structure in a neighbourhood of the knot. Finally, we discuss computational implementation of the formulae derived, with C code provided, and give illustrations for how the solid angle may be used to give explicit constructions of knotted scroll waves in excitable media and knotted director fields around disclination lines in nematic liquid crystals. 
    \end{abstract}


    %\keywords{knotted fields, solid angle, geometry}
    %\submitto{\jpa}
    %\maketitle

    \section{Introduction}
    \label{sec:intro}

    Knotted fields are three-dimensional textures of continuous media that encode in their structure a knotted curve, filament or family of field lines. Originating in Lord Kelvin's speculations of atomic structure as knotted vortices in the aether~\cite{Thomson1867}, they have since been experimentally realised in nodal lines of optical beams~\cite{Dennis2010}, disclinations in nematic liquid crystals~\cite{Tkalec2011,Tasinkevych2014,Copar2015}, spinor Bose-Einstein condensates and fluid vortices~\cite{Kleckner2013}. Concurrently, theoretical studies continue to flourish in classical field theory~\cite{Sutcliffe2007}, electromagnetism~\cite{Kedia2013,Arrayas2017}, superfluids \cite{Kleckner2016} and excitable media~\cite{Maucher2016,Maucher2017,Maucher2018}.

    Central to theoretical advances are explicit constructions for knotted fields exhibiting different knot types, or other pertinent physical properties, such as helicity in fluid flows. Constructions for knots in electromagnetic fields have centred around the Hopf map and rational map generalisations of it, shear-free null congruences and twistor methods~\cite{Arrayas2017,Ranada1992,Kedia2013,Kedia2018}. The simplest constructions yield torus knots and links and the majority of constructions have focused on this family, together with seeking to control the helicity of the field~\cite{Kedia2018}, or its dynamics~\cite{Irvine2010}. The same rational map constructions also give knotted solutions in other field theories, such as the Skyrme-Faddeev model~\cite{Sutcliffe2007,Battye1998}. These methods satisfy the dynamical equations of motion directly and are geometrically special by construction, providing powerful tools for describing the full knotted field and its properties. 

    A separate approach has been developed to create nodal lines in optical beams that encodes the knot as the zero locus of a complex polynomial~\cite{Dennis2010}. From these fields initial conditions can be generated for paraxial wave equations with the subsequent evolution giving a beam containing the encoded knot. Again, the simplest constructions are for torus knots (captured by the polynomials $z_1^p+z_2^q$) but the method can be applied for any geometric braid~\cite{Bode2017,Dennis2017}. The argument of such a complex polynomial gives a phase field that winds around the knotted nodal line and can be used to initialise phase vortices, or as an angle orienting the director field of a liquid crystal with the nodal line then appearing as a disclination~\cite{Machon2014}. In common with the constructions for electromagnetic knots, this approach encodes the knot implicitly rather than explicitly in that its location and geometry derives from the polynomial rather than being given {\sl a priori}. 

    A canonical construction for a phase field associated to any knotted curve $K$, that depends only on the curve and represents a knotted field on its complement is given by the solid angle $\omega(\bf{x})$ subtended by $K$ at each point in space. This construction of knotted fields goes back to Maxwell~\cite{Maxwell2}, since the solid angle is proportional to the magnetostatic potential of a current carrying wire, and in all likelihood represents the earliest explicit construction for a knotted field. 
    If we imagine $K$ to be a wire carrying unit current then Maxwell's equations state that it generates, in its complement, a magnetic field that is irrotational, so that locally it is the gradient of a potential. Amp\`ere's law shows this potential to be globally multi-valued (increasing by $\mu_0$ upon traversing any closed loop encircling the wire): the solid angle is the magnetostatic potential normalised to be $4\pi$ cyclic, {\sl i.e.} it takes values in $\mathbb{R}/4\pi\mathbb{Z}\cong S^1$. This description makes clear that solid angle is naturally defined for an oriented curve $K$, the orientation being provided by the current flow. Since magnetic fields are divergence free, the solid angle is a harmonic function, and this, together with the $4\pi$ circulation, may be taken as an alternative definition. Knotted fields constructed out of it satisfy physical differential equations (Laplace's equation), but in contrast to other methods are more direct and explicit in their construction, so that there is no special focus on torus knots, geometric braids or any other particular class of knots. 

 Construction of the magnetostatic potential via numerical integration of the magnetic field about $K$ has recently been used to initialise knotted fields in superfluids and excitable media~\cite{Kleckner2016,Maucher2016}. However, very little in the way of a systematic treatment of solid angle and its geometric content has been given since Maxwell's own presentation in his {\sl Treatise on Electricity and Magnetism}~\cite{Maxwell2}. Maxwell devotes articles~417-422 of Ref.~\cite{Maxwell2} to an extended discussion of solid angle, its properties and geometric meaning, as well as methods for calculating it. He gives three methods, in addition to \eref{eq:SolidAngle1}: a direct calculation; a method given ``for the sake of geometrical propriety''; and his preferred method which involves calculating the work done in transporting a unit magnetic pole to the point ${\bf x}$. Through the latter he (independently) derives the Gauss linking integral~\cite{Ricca2011}. 

 Typically, solid angle is described with the help of an orientable surface $\Sigma$ spanning $K$: $\omega({\bf x})$ is then the area that this surface projects to on the unit sphere centred on $\bf x$, and is given explicitly by the formula~\cite{Saffman1993} (which Maxwell attributes to Gauss~\cite[Art.~409]{Maxwell2})
    \begin{equation}
    \omega({\bf x}) = \int_{\Sigma} \frac{({\bf x}-{\bf y})}{|{\bf x}-{\bf y}|^3}\cdot \mathrm{d}{\bf S} ,
    \label{eq:SolidAngle1}
    \end{equation}
    where ${\bf y}$ varies over $\Sigma$. While this description hides the fact that solid angle depends only on $K$, it provides the main geometric interpretation for solid angle and establishes close connections to projective and spherical geometry, particularly to spherical curves and areas. Solid angle, then, is a naturally geometric object dependent only on $K$, which involves an interplay between the geometry of $K$ itself, and that of the spherical curve to which $K$ projects. As such it belongs firmly to the domain of the differential geometry of curves. Yet its relationship to curve geometry is only partially developed, limited to how the local geometry influences the local structure of the magnetic field in the curve's normal plane~\cite{Saffman1993,Moore1972,Ricca1994}. A related question is that of an `optimal' method of computing $\omega$, both from a theoretical and computational standpoint. Both methods mentioned above suffer deficiencies. In the first, an unnecessary intermediate, the magnetic field, is computed before $\omega$. In the second, an arbitrary surface spanning $K$ must be provided, of which $\omega$ is independent --- this is especially inconvenient from a numerical standpoint. We desire a convenient direct expression for $\omega$, dependent only on $K$. 

    In this paper, we show that Maxwell's three methods, extended where appropriate to knotted curves, may all be considered as applications of a single curve homotopy formula. In doing so, we shall arrive at several distinct formulae for computing $\omega$ directly from $K$ and make connections between solid angle and modern results on the geometry of spherical curves~\cite{Levi1994,Arnold1995}, as well as discussing close connections between the asymptotic structure of $\omega$ and the writhe of $K$~\cite{Fuller1978,Dennis2005}. With these formulae in place, we offer a geometric description of the local properties of $\omega$ in a tubular neighbourhood of $K$, considering both the structure in the normal plane and as one moves along the knot. Our description, which begins directly at the spherical geometry of the projected curve, complements existing results on the local structure of the magnetic field, and reveals a previously unseen connection between the local structure of $\omega$ and the `writhe framing' of Ref.~\cite{Dennis2005}. 
    Our results give several formulae for the direct computation of $\omega$ from $K$, of practical value when initialising simulations of knotted fields. We discuss solutions to the main difficulties in their numerical implementation, and end with a brief description of applications to the initialisation of scroll waves in excitable media and knotted textures in nematics. Implementations in C of the methods described are given at \verb=github.com/garethalexander=.

The extension of the construction of solid angle to the case where $K$ is a link is straightforward: by the linearity of electromagnetism the solid angle for a link is simply the sum (mod $4\pi$) of the solid angles corresponding to each of the link components. For this reason, we restrict the majority of our discussion to knots, and discuss the few subtleties which come with extension to links in a brief dedicated section.

    \section{The homotopy formula for solid angle} 
    \label{sec:CurveIsotopies}

    \begin{figure}[t]
    \centering	
    \includegraphics[width=\textwidth]{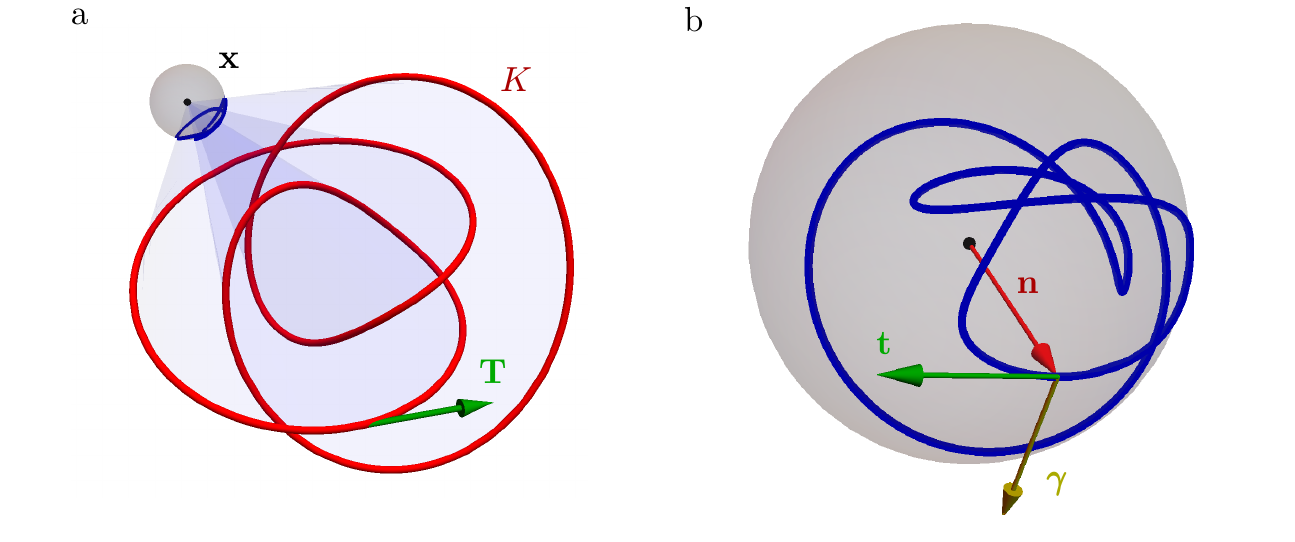}
    \caption{a) An oriented knot $K$ with tangent vector $\bf T$ (here the $4_1$) projects onto a unit observation sphere about a point $\bf x$, giving the spherical curve shown in blue. b) The projection of $K$ onto the observation sphere gives an immersed spherical curve $\bf n$, with self-intersections in correspondence with the crossings of the knot as seen from $\bf x$. A unit tangent $\bf t$ for $\bf n$ is induced by the orientation of $K$, and we select normal $\boldsymbol{\gamma} := \bf n \times \bf t$.}
    \label{fig:Knot} 
    \end{figure}

    At each point ${\bf x}$ of the knot complement the projection of $K$ onto the unit sphere centred on ${\bf x}$, which we shall call the observation sphere, traces out a curve ${\bf n}:=\frac{{\bf y}-{\bf x}}{|{\bf y}-{\bf x}|}$, ${\bf y}\in K$, as shown in Fig.~\ref{fig:Knot}. This projected curve has points of self-intersection in correspondence with the crossings of the knot as seen from ${\bf x}$. Upon varying ${\bf x}$ there will be particular viewing points where the number of visible crossings changes and at those points ${\bf n}$ also has cusps. In all cases \eref{eq:SolidAngle1} expresses that the solid angle at ${\bf x}$ is the area bound by the projected curve ${\bf n}$ on the observation sphere; indeed, Maxwell states this as the definition of the solid angle. 

    Maxwell's first method of computing $\omega({\bf x})$ is to choose arbitrary spherical coordinates $(\theta,\phi)$ on the observation sphere, and integrate the projected area directly~\cite[Art.~417]{Maxwell2}:
    \begin{equation}
    \omega({\bf x}) = \int (1 - \cos \theta)~\mathrm{d}\phi .
    \label{eq:MaxSphere}
    \end{equation}
    If we denote by ${\bf n}_{\infty}$ the (arbitrarily chosen) polar direction $\theta=0$, then \eref{eq:MaxSphere} can be expressed in vector notation as 
    \begin{equation}
    \omega({\bf x}) = \int \frac{{\bf n}_{\infty} \times {\bf n} }{1+{\bf n}_\infty \cdot {\bf n}} \cdot \mathrm{d}{\bf n} ,
    \label{eq:OurSolidAngle1}
    \end{equation}
    a formula that has been rediscovered a number of times~\cite{Asvestas1985,Dangskul2015,Borodzik2017}. We remark that if we interpret \eref{eq:OurSolidAngle1} as an integral over $K$ rather than its projection on the observation sphere, the integrand is the vector potential for a magnetic monopole placed at ${\bf x}$, with $-{\bf n}_{\infty}$ corresponding to the choice of Dirac string. Indeed, expressing it in the spherical coordinates of \eref{eq:MaxSphere} we recover the vector potential of Ref.~\cite{Dirac1931}
\begin{equation}
\frac{{\bf n}_{\infty} \times {\bf n} }{1+{\bf n}_\infty \cdot {\bf n}} \cdot \frac{1}{| \bf {y - x} |} = \frac{\sin{\theta}}{r(1+\cos{\theta})}\boldsymbol{\hat{\phi}}, 
    \end{equation}
    where $r = |{\bf y -x}|$. Maxwell gives this formula explicitly in Cartesian coordinates and remarks on the role of the string (``axis'') in evaluating the integral. 

    Maxwell does not advocate the use of \eref{eq:MaxSphere}, other than for computational convenience, writing that it ``involves a choice of axes which is to some extent arbitrary, and it does not depend solely on the closed curve''~\cite[Art.~418]{Maxwell2}. We shall discuss his second method in \S\ref{sec:GaussBonnet}, but his preferred method is his third ``as it employs no constructions which do not flow from the physical data of the problem''~\cite[Art.~419]{Maxwell2}: viewing $\omega$ as the magnetostatic potential of $K$, it may be built by measuring the change $\Delta\omega$ as we transport a unit magnetic pole along an arbitrary path from a reference location to ${\bf x}$, or equivalently by fixing ${\bf x}$ and oppositely transporting $K$. Maxwell gives a formula for $\Delta\omega$ under this transport in terms of a double integral over the path and $K$, by summing the areas of the infinitesimal parallelograms swept out by line elements of $K$. 

    This approach shifts the focus from calculating the solid angle directly to calculating the change induced by a translation of the knot along some path. It is a small step to extend this to give a formula for the change associated to a general homotopy of $K$, in which the shape of $K$ may vary. Of course, $\Delta \omega$ does not depend on the precise form of this homotopy, which allows it to be calculated using a standardised method, for instance by connecting corresponding points of the initial ($K_0$) and final ($K_1$) curves with straight lines, {\sl i.e.} $K_t = (1-t) K_0 + t K_1$, $t\in [0,1]$. This homotopy induces one on the observation sphere, which we denote ${\bf n}_t$, with the straight lines along which the points of $K$ move projecting to geodesic arcs connecting ${\bf n}_0$ and ${\bf n}_1$. The change in solid angle is the area swept out by this mesh of geodesic arcs. 

    Consider the contribution to the area of the geodesics connecting a small segment of the two curves: By Archimedes' theorem on the equality of the area of the sphere and its circumscribed cylinder this is equal to the product of the distance $|{\bf n}_0-{\bf n}_1|$ between the two endpoints of the geodesic arc and the angle swept out by its midpoint $({\bf n}_0+{\bf n}_1)/|{\bf n}_0+{\bf n}_1|$. The difference in solid angle is therefore 
    \begin{eqnarray}
    \omega({\bf x}; K_1) - \omega({\bf x}; K_0) &= \int ({\bf n}_0 - {\bf n}_1) \times \frac{{\bf n}_0 + {\bf n}_1}{|{\bf n}_0 + {\bf n}_1|} \cdot \mathrm{d} \frac{{\bf n}_0 + {\bf n}_1}{|{\bf n}_0 + {\bf n}_1|} \quad  \nonumber \\ 
    &= \int \frac{{\bf n}_0 \times {\bf n}_1 \cdot ( \mathrm{d}{\bf n}_0 + \mathrm{d}{\bf n}_1 )}{1+{\bf n}_0 \cdot {\bf n}_1} \quad \textrm{mod } 4\pi.  \label{eq:Isotopy}
    \end{eqnarray}
    This is the basic homotopy formula for solid angle, applicable to an arbitrary deformation of $K$. Both Maxwell's first and third methods of computing $\omega$ can be seen as applications of \eref{eq:Isotopy} --- we recover \eref{eq:OurSolidAngle1} by letting $K_0$ recede asymptotically far from $\bf x$, so that ${\bf n}_0$ is a single point ${\bf n}_{\infty}$ on the observation sphere and $\omega({\bf x}; K_0)=0$ mod $4\pi$. In \S\ref{sec:GaussBonnet} we shall use a homotopy of $K$ along its tangent developable surface to demonstrate that his second method also follows directly from the homotopy formula. 

    The integral in \eref{eq:Isotopy} is not defined when $\bf x$ lies on the surface swept out by $K_t$, which we refer to as the surface of discontinuity --- as an example, in \eref{eq:OurSolidAngle1} this surface is formed by translating $K$ to infinity along ${\bf n}_\infty$. The line of $K_t$ passing through $\bf x$ connects antipodal points of the observation sphere, ${\bf n}_0 \cdot {\bf n}_1= -1$, and this line does not project to a unique geodesic arc connecting these endpoints. Instead there is a whole family of equivalent connecting geodesics, which cover the sphere once. As $\bf x$ crosses the surface of discontinuity, the geodesic parameterisation of the antipodal sections of ${\bf n}_0$ and ${\bf n}_1$ jumps from one side of the observation sphere to the other, giving a $4 \pi$ jump in \eref{eq:Isotopy}.

    We note that~\eref{eq:Isotopy} has the same form as the formula given by Fuller for the difference in writhe of two curves~\cite{Fuller1978}. This is because for each fixed point ${\bf x}$ (not on $K_t$ for any $t$) the difference in solid angle is expressible as an area between two spherical curves, as arises for the difference in writhe. This is the first of several relations between the solid angle function for a curve and its writhe, which help to convey its geometric content.

    \section{Maxwell's geometric formula, dual curves and homotopies along tangent developable surfaces }
    \label{sec:GaussBonnet}
    \begin{figure}
    \centering
    \includegraphics{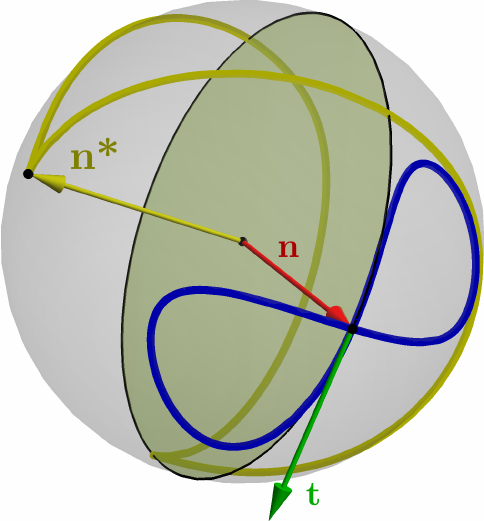}
    \caption{A spherical knot projection $\bf n$ (blue curve, here a typical projection of a twisted unknot) induces a dual spherical curve ${\bf n}^* := {\bf t} \times {\bf n}$ (yellow curve). Maxwell proposes the construction of ${\bf n}^*$ by allowing a unit circle (yellow disk) to roll without slipping around $\bf n$ such that its plane of contact is tangent to $\bf n$. A unit vector perpendicular to this circle (yellow arrow) then traces ${\bf n}^*$. As shown in \eref{eq:dn*ds}, zeros of geodesic curvature in $\bf n$ correspond to cusps in $\bf n^*$ (marked points). More pictures of this construction may be found in Refs. \cite{Levi1994,Arnold1995}.   }
    \label{fig:DualCurve} 
    \end{figure}
    Maxwell's objection to \eref{eq:MaxSphere} is that it involves an arbitrary choice of spherical coordinates on the observation sphere, and for this reason he states a construction in which no such choice is made \cite[Art.~418]{Maxwell2}. Let a unit circle roll without slipping around ${\bf n}$ such that its plane of contact is tangent to $\bf n$, as shown in Fig.~\ref{fig:DualCurve}. Then a unit vector perpendicular to this circle traces a second curve on the observation sphere, called the dual curve ${\bf n}^*$. Denote the length of ${\bf n}^*$ by $\sigma$. Maxwell states that the solid angle is given by 
    \begin{equation}
    \omega({\bf x}) = 2\pi - \sigma ,
    \label{eq:DualCurve}
    \end{equation}
    a result he simply describes as a ``well-known theorem''. This result is in fact equivalent to the Gauss-Bonnet formula~\cite{Lee1996}, an identification that has been rediscovered at least twice~\cite{Levi1994,Arnold1995}. In the form stated by Maxwell,~\eref{eq:DualCurve} is only correct if ${\bf n}$ is a simple curve without points of inflection, but it is true in much greater generality~\cite{Arnold1995}. As a more general version is essential for application to generic knot projections, we give a self-contained elementary proof, applicable to any smoothly immersed spherical curve. 

    \subsection{A dual curve theorem for self-intersecting curves}
    \label{subsec:GaussBonnetSelfIntersecting}
    \begin{figure}
    \centering
    \includegraphics[width=\textwidth]{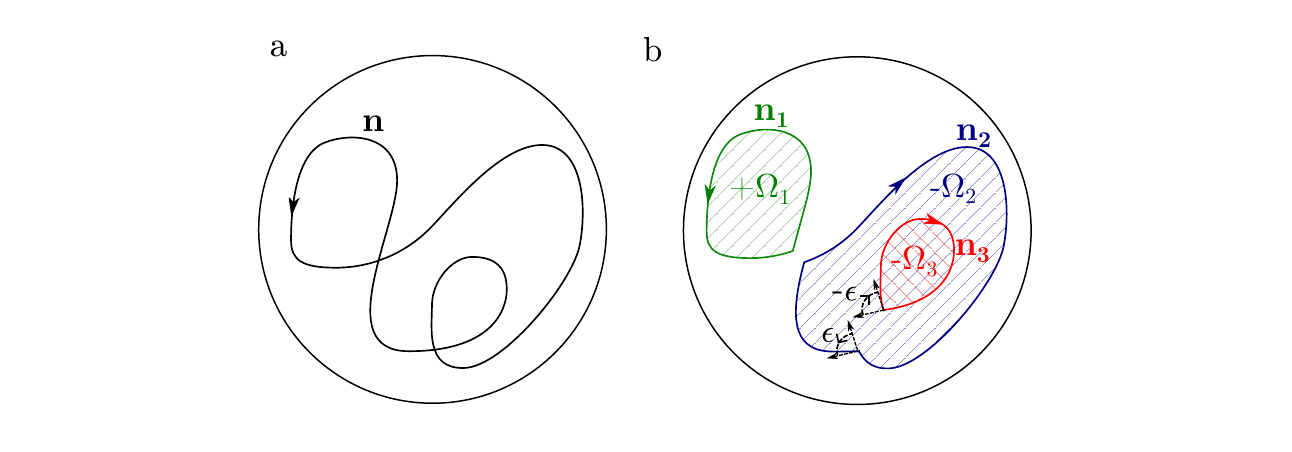}
    \caption{The spherical knot projection $\bf n$ in a) may be decomposed via the Seifert algorithm into Seifert circles ${\bf n}_i$, shown in b). These circles bound regions $\Omega_i$, signed according to the orientation of their boundary (coloured hatching). At self-intersection points of $\bf n$, the resulting circles have corners, with exterior angles $\epsilon_{ij}$ (shown for one such corner).}
    \label{fig:GaussBonnet} 
    \end{figure}

    We begin by relating the area swept out by ${\bf n}$ to its integrated geodesic curvature by using the Gauss-Bonnet formula. ${\bf n}$ has a canonical tangent vector induced from the orientation of $K$, denoted $\bf t$, and we choose for it a normal vector $\boldsymbol \gamma := \bf n \times \bf t$, as shown in Fig.~\ref{fig:Knot}b). (Note that in the special case that ${\bf n}$ is a simple curve it bounds two regions on the sphere, but is only correctly oriented as the boundary of one of them. ${\boldsymbol \gamma}$ points inwards to this region.) We perform a Seifert decomposition~\cite{Adams2004} of ${\bf n}$. This entails resolving each crossing in a manner that preserves the orientation of the curve and results in its separation into a collection of Seifert circles ${\bf n}_i$, as shown in Fig.~\ref{fig:GaussBonnet}. Each circle is a simple curve and bounds a region $\Omega_i$. At self-intersections of ${\bf n}$ the Seifert circles have corners, with exterior angles $\epsilon_{ij}$. Now, for each circle, the Gauss-Bonnet formula tells us
    \begin{equation}
    \int_{\Omega_i} \mathrm{d} A  = 2 \pi - \int_{{\bf n}_i} k_{\boldsymbol \gamma} \mathrm{d} s - \sum_j \epsilon_{ij} ,
    \label{eq:GaussBonnetSeifert}
    \end{equation}
    where $k_{\boldsymbol \gamma} = \frac{\rmd {\bf t}}{\rmd s} \cdot {\boldsymbol \gamma}$ is the signed geodesic curvature of the boundary. Summing over all Seifert circles, the left-hand-side gives $\omega({\bf x})$ mod $4\pi$; on the right-hand-side the exterior angles cancel pairwise, and we pick up a contribution of $2\pi S$, where $S$ is the number of Seifert circles, in addition to the total integrated (signed) geodesic curvature. The number of Seifert circles is equal to $\chi + D$, where $\chi$ is the Euler characteristic of the surface constructed by the Seifert algorithm and $D$ is the number of double points (self-intersections)~\cite{Adams2004,Lickorish1997}. For a knot the Euler characteristic of any Seifert surface is odd, so that $S = D+1$ mod $2$. Thus we have
    \begin{equation}
    \omega({\bf x})  = 2 \pi (D + 1) - \int_{{\bf n}} k_{\boldsymbol \gamma} \mathrm{d}s\quad\textrm{mod}\; 4\pi .
    \label{eq:GaussBonnetImmersed}
    \end{equation} 
    We remark that the quantity $D+1$ mod $2$ is the spherical equivalent of the rotation number of a planar self-intersecting curve, sometimes termed its parity \cite{Whitney1937,Phillips1966,Solomon1996}. The $2 \pi$ in the Gauss-Bonnet formula arises as the rotation number of a simple curve, and the appearance of the parity here is thus a natural extension to the self-intersecting case. 

    The integrated geodesic curvature is equal to the (signed) length of the dual curve ${\bf n}^* := -{\boldsymbol \gamma} = {\bf t} \times {\bf n}$ \cite{Levi1994,Arnold1995}(Fig.~\ref{fig:DualCurve}). To see this, consider how ${\bf n}^*$ varies with arc length along ${\bf n}$:
    \begin{equation}
    \frac{\mathrm{d}{\bf n}^*}{\mathrm{d}s} = \frac{\mathrm{d}\,}{\mathrm{d}s}({\bf t} \times {\bf n}) = \frac{\mathrm{d}{\bf t}}{\mathrm{d}s} \times {\bf n}= k_{{\boldsymbol \gamma}} {\bf t} .
    \label{eq:dn*ds}
    \end{equation}
    ${\bf t}$ is tangent to ${\bf n}^*$, but its orientation alternates across zeros of $k_{\boldsymbol \gamma}$, which correspond to cusps in ${\bf n}^*$. Defining $ds^* = k_{\boldsymbol \gamma}ds$, we obtain $\mathrm{d}{\bf n}^*/\mathrm{d}s^* = {\bf t}$, and see that $ds^*$ should be interpreted as a signed length element, the sign being given by that of $k_{\boldsymbol \gamma}$. Thus we arrive at
    \begin{equation}
    \omega({\bf x}) =  2 \pi (D+1) - \int_{{\bf n}^*} \mathrm{d} s^* \quad \mathrm{mod}\; 4\pi . 
    \label{eq:DualCurveImmersed}
    \end{equation}
    For a simple curve without inflection points $D=0$ and the sign of $\mathrm{d}s^*$ never alternates, so its integral gives $\sigma$ and we recover \eref{eq:DualCurve}. By contrast, for $\bf {n}$ as shown in Fig.~\ref{fig:DualCurve} $D=1$ and we have two zeros of geodesic curvature, which divide $\bf{n}^*$ into two segments separated by cusps with $ds^*$ switching sign between them. Applying \eref{eq:DualCurveImmersed} to this example gives the expected result $\omega =0$; applying \eref{eq:DualCurve} does not. Eq. \eref{eq:DualCurveImmersed} thus generalises Maxwell's ``well known theorem'' \eref{eq:DualCurve} to the case of a smoothly immersed curve, and in particular to any generic spherical knot projection. 

    \subsection{The pullback to $K$ and a homotopy along the tangent developable surface}
    \label{subsec:Pullback}

    As a result on the structure of spherical areas,~\eref{eq:DualCurveImmersed} is valid for any spherical curve. However, we have in mind the case where one arises as the projection of the knot $K$. Using this projection we now pull each term in~\eref{eq:DualCurveImmersed} back to $K$. This facilitates a reinterpretation in terms of the geometry of $K$, as well as a novel method of deriving it using~\eref{eq:Isotopy}.

    We begin by constructing a natural `projective' framing for $K$, dependent on $\bf x$, with which we will express $D$ in~\eref{eq:DualCurveImmersed} as a self-linking number. To construct this framing, extend the lines of sight from ${\bf x}$ along ${\bf n}$ until they meet $K$. These lines project to vectors normal to $K$, which are non-zero provided ${\bf n} \cdot {\bf T} \neq \pm 1$ where ${\bf T}$ is the unit tangent vector to $K$, in other words provided there are no cusps in ${\bf n}$ on the observation sphere. The number of double points seen from $\bf x$ mod $2$ is equal to the self-linking number of $K$ given this projective framing, $\mathrm{SL}(K,{\bf x})$, also mod $2$. The mod $2$ counting gives an ambiguity in the sign of the identification of $D$ with $\mathrm{SL}(K,{\bf x})$ which will lead to two distinct re-writings of~\eref{eq:DualCurveImmersed}, and so we shall keep the sign explicit in the following. 

    Using C\u{a}lug\u{a}reanu's theorem~\cite{Calugareanu1959,Calugareanu1961}, $ \mathrm{SL}(K,{\bf x}) = \mathrm{Tw}(K,{\bf x}) +\mathrm{Wr}(K)$, we now write $ \mathrm{SL}(K,{\bf x})$ in terms of the writhe of $K$ and the twist of the projective framing, which is directly computed to be 
    \begin{equation}
    \mathrm{Tw}(K,{\bf x}) = \frac{1}{2\pi} \int_K  \frac{ ( {\bf n} \cdot {\bf T} ) ({\bf n} \cdot {\bf T} \times \mathrm{d}{\bf T}) }{1 - ({\bf n} \cdot {\bf T})^2} . 
    \label{eq:Twist}
    \end{equation}
    Substituting this expression for $\mathrm{SL}(K,{\bf x})$ into \eref{eq:DualCurveImmersed} with the sign ambiguity discussed above, and combining with the pullback of the dual curve length,
    \begin{equation}
    \int_{{\bf n ^*}} \mathrm{d} s^* = \int_{{\bf n}} k_{\bf \boldsymbol \gamma } \mathrm{d} s = \int_K  \frac{ {\bf n} \cdot {\bf T} \times \mathrm{d}{\bf T} }{1 - ({\bf n} \cdot {\bf T})^2},
    \label{eq:geodesicpullback1}
    \end{equation}
    we arrive at
    \begin{equation}
    \omega({\bf x}) = 2 \pi(1 \pm \mathrm{Wr}(K)) - \int_K  \frac{ {\bf n} \cdot {\bf T} \times \mathrm{d}{\bf T} }{1 \pm {\bf n} \cdot {\bf T}}  \quad \mathrm{mod}\; 4\pi .
    \label{eq:OurSolidAngle2}
    \end{equation}    
    \begin{figure}[t]
    \begin{centering}
    \includegraphics[ width = 0.4\textwidth ]{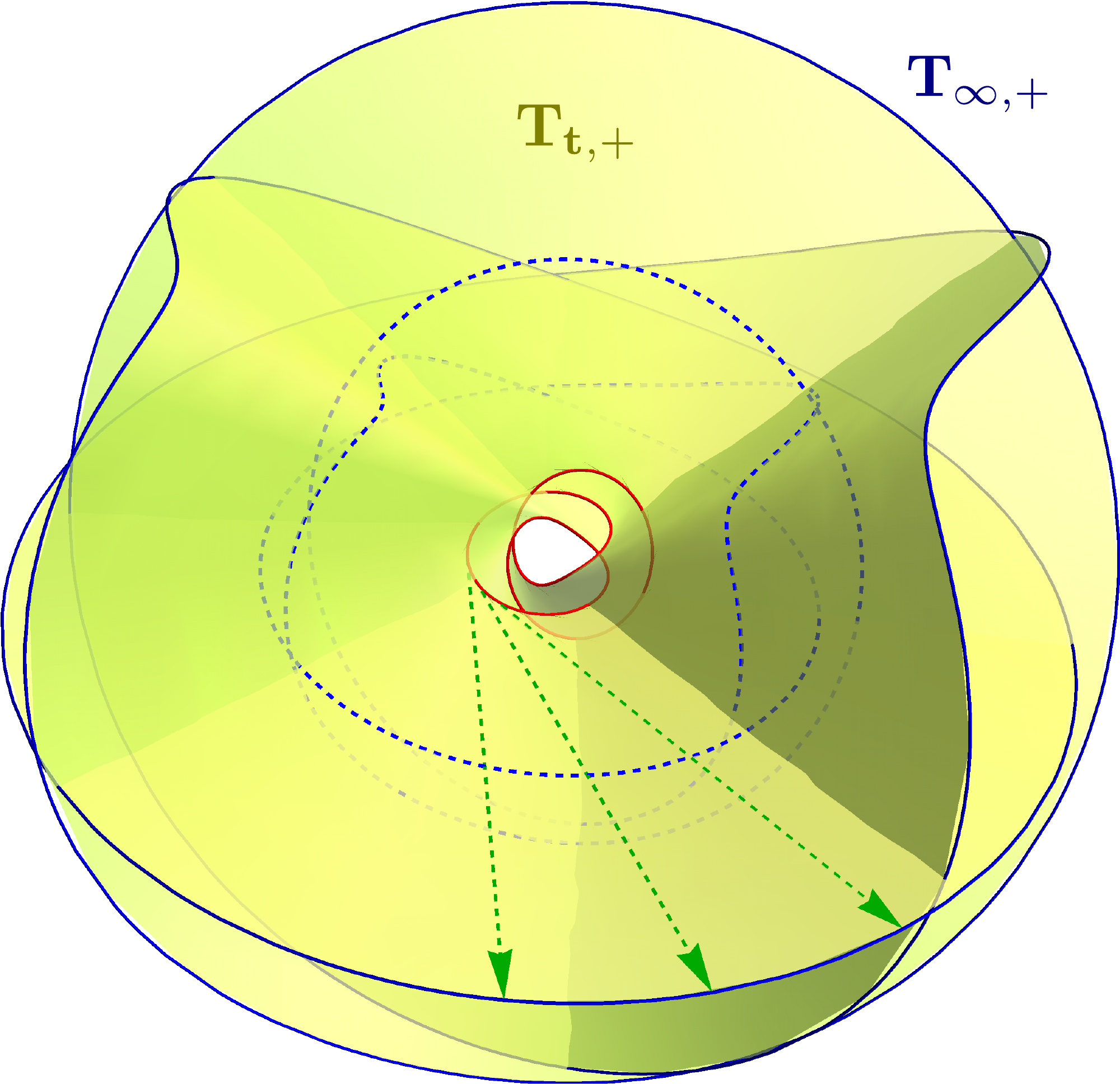}
    \caption{The forward tangent developable surface ${\bf{T}}_{t,+}$ (yellow surface) for the knot $K$ in Fig.~\ref{fig:Knot} (red curve), constructed by extending half-lines along tangents from $K$ (green, dashed). The intersection of the surface with a sphere of asymptotically large radius gives a scaled copy of the tangent indicatrix to $K$, ${\bf T}_{\infty,+}$ (blue). The half-lines comprising ${\bf T}_{t,+}$ define a straight line homotopy between $K$ and ${\bf T}_{\infty,+}$, from which the blue, dashed curve is taken.}
    \label{fig:TangentDevelopable}
    \end{centering}
    \end{figure}
    This formula for the solid angle depends only on $K$ and data canonically associated to it, with the only ambiguity being a choice of sign. The appearance of the writhe in \eref{eq:OurSolidAngle2} reveals this geometric property of curves to be closely connected to the solid angle. We shall return to the sign ambiguity in a moment --- for now, let us select the plus sign. 
    
    Instead of taking \eref{eq:DualCurve} as our starting point, we now demonstrate how \eref{eq:OurSolidAngle2} may be derived directly from the curve homotopy formula \eref{eq:Isotopy}. To construct the appropriate homotopy, extend half-lines from $K$ along its tangents ${\bf T}$, sweeping out a surface in space known as the forward tangent developable surface of $K$, which we denote ${\bf{T}}_{t,+}:={\bf y} + t {\bf T}$, $t\in [0,\infty)$~\cite{Eisenhart} --- an example of this surface is shown in Fig.~\ref{fig:TangentDevelopable}. Consider the intersection of this surface with a sphere of asymptotically large radius. The curve ${\bf T}_{\infty, +}$ given by this intersection is simply the spherical image of $\bf T$, known as the forward tangent indicatrix of $K$ ~\cite{Eisenhart}, scaled to the sphere radius. Our desired homotopy is between $\bf T_{\infty,+}$ and $K$, and is defined by the half-lines comprising ${\bf{T}}_{t,+}$. As $\bf T_{\infty,+}$ is asymptotically far from ${\bf x}$, its projection on to the observation sphere simply reproduces the tangent indicatrix. Using the fact that ${\bf n} \times {\bf T} \cdot \mathrm{d}{\bf n} = 0$, we see that the integral in \eref{eq:OurSolidAngle2} is a second special case of \eref{eq:Isotopy}, with $K_0 = {\bf T}_{\infty,+}$, $K_1 =K$, and the area swept out on the observation sphere lying between the forward tangent indicatrix and $\bf n$. 

    This argument also identifies $2\pi(1+\mathrm{Wr}(K))$ as the solid angle of $\bf T_{\infty,+}$. We may obtain an integral formula for this area by considering the asymptotics of \eref{eq:OurSolidAngle2}, allowing ${\bf x}$ to recede far from $K$ along $-{\bf n}_{\infty}$ so that $\omega({\bf x}) \rightarrow 0$. Doing so yields
    \begin{equation}
    \int_K  \frac{ {\bf n}_\infty \cdot {\bf T} \times \mathrm{d}{\bf T} }{1 + {\bf n}_\infty \cdot {\bf T}} = 2 \pi(1 + \mathrm{Wr}(K))  \quad \textrm{mod}\; 4\pi , 
    \label{eq:AsymptoticSolidAngle2}
    \end{equation}
    however, as this integral is the area bound by the tangent indicatrix on the unit sphere, the identification is simply a recovery of Fuller's writhe mod $2$ formula~\cite{Fuller1978}. In the context of curve homotopies, we may interpret \eref{eq:AsymptoticSolidAngle2} as giving the change in solid angle for a homotopy in which $\bf T_{\infty,+}$ shrinks to a point (that projects to ${\bf n}_\infty$ on the observation sphere). Eq. \eref{eq:OurSolidAngle2} may then be thought of as a combination of two homotopies: the first from an arbitrary point to $\bf T_{\infty,+}$, and the second from $\bf T_{\infty,+}$ to $K$. By contrast, \eref{eq:OurSolidAngle1} combines these two homotopies into one. Returning to the sign choice made above, we now see that choosing a minus sign would give a version of \eref{eq:OurSolidAngle2} corresponding to a homotopy along the backward tangent developable surface ${\bf{T}}_{t,-}:={\bf y} - t {\bf T}$, $t\in [0,\infty)$, between $K$ and the backward tangent indicatrix $\bf T_{\infty,-}$. That aside, the geometric interpretation remains the same. We note briefly that the tangent indicatrix is not the only spherical curve canonically associated with $K$ which might be used to define a homotopy; we might also consider the normal and binormal indicatrices. In these cases, however, neither triple product in \eref{eq:Isotopy} vanishes, as occurred in \eref{eq:OurSolidAngle2}, and so the resulting formulae are less simple. 

     With the choice of plus (minus) sign in \eref{eq:OurSolidAngle2}, the surface of discontinuity discussed in \S\ref{sec:CurveIsotopies} is given by ${\bf T}_{t,+}$ (${\bf T}_{t,-}$). Jumps are also present in~\eref{eq:GaussBonnetImmersed} and~\eref{eq:DualCurveImmersed}, however they occur on both halves of the tangent developable surface ${\bf T}_{t,+}\cup {\bf T}_{t,-}$ and the overall $4 \pi$ jumps are composed of each individual term in the equations jumping by $2 \pi$. To convince ourselves of this fact, consider the behaviour of \eref{eq:DualCurveImmersed} as ${\bf x}$ passes across ${\bf T}_{t,+}\cup {\bf T}_{t,-}$. ${\bf n}$ undergoes a Reidemeister 1 move, during which $D$ jumps by 1. The segment of ${\bf n}^*$ corresponding to the Reidemeister move in ${\bf n}$ begins and ends at antipodal points on the sphere. By removing the loop in ${\bf n}$, we create two inflection points. Recalling that the sign of $ds^*$ alternates between these inflections, we pick up a change in signed length of $2 \pi$.

    \section{The structure of $\omega$}
    \label{sec:LocalStructure}

\begin{figure}[t]
    \begin{centering}
    \includegraphics[clip = true, trim = 120 380 120 0, width = \textwidth ]{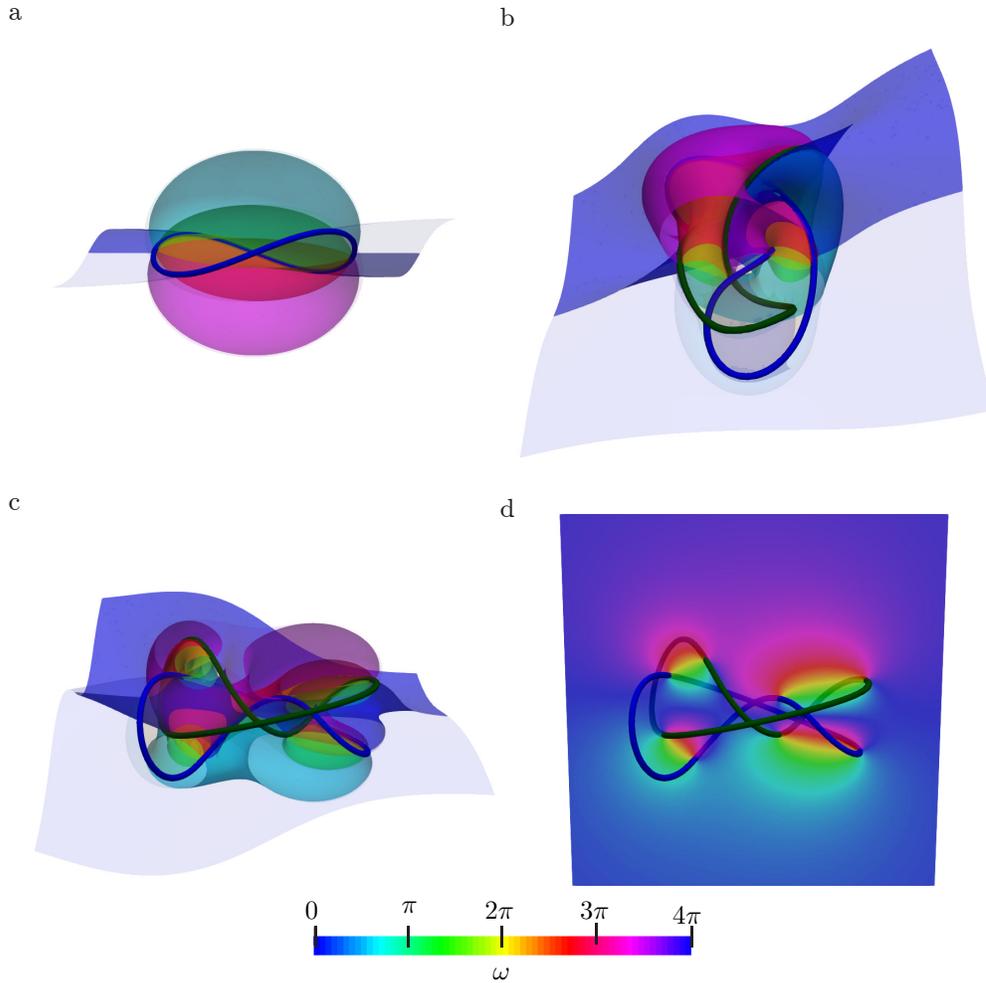}
    \caption{The structure of $\omega$ around a knotted curve, generated with the method of \S\ref{sec:NumericalImplementation}. a--c) show level sets of $\omega$ of spacing $\frac{\pi}{2}$, each of which forms a Seifert Surface for the knot with opacities on the near sides of the images reduced to reveal the inner structure of $\omega$. a) A twisted unknot. b,~c) The Whitehead link (components in blue, green) from two viewing directions. d) A slice through the Whitehead link from the same direction as c). The local structure of $\omega$ about the knot is especially clear in d) --- $\omega$ winds by $4 \pi$, and as we move away from the knot, curvature induced corrections cause the level sets of $\omega$ to bunch along the curve normal, as discussed in \S\ref{sec:LocalStructure}.}
    \label{fig:SolidAngle}
    \end{centering}
    \end{figure}

The level sets of $\omega$, for regular values, form a family of Seifert surfaces with common boundary $K$. Figure~\ref{fig:SolidAngle} shows this global structure for a twisted unknot and a Whitehead link. 
The topology of the level sets changes at critical points of $\omega$, where generically the local structure is a cone point $\pm(x^2+y^2-2z^2)$ with Morse index $1$ or $2$. As the solid angle is a harmonic function, critical points of Morse index $0$ or $3$ are forbidden by the maximum principle. For knots and links that are fibred~\cite{Stallings1978} it is possible for the solid angle to have no critical points at all; indeed this is the case for both the unknot and Whitehead link shown in Fig.~\ref{fig:SolidAngle}. The general relationship between the shape and geometry of a knot or link and critical points of the solid angle is a fascinating open problem. 

It is of particular interest to characterise $\omega$ in a tubular neighbourhood of $K$, so that we may modify it when initialising simulations using $\omega$. This control is useful when the local structure of the field around a vortex affects its dynamics, as for example in helicity in fluids~\cite{Moffatt1992} or the twist of scroll waves in the FitzHugh-Nagumo model~\cite{Winfree1984,Maucher2018}. This local structure has longitudinal and transverse parts: the level sets of $\omega$ rotate as one traverses $K$, and in a plane normal to $K$ corrections due to local curvature and torsion arise, analogous to those studied for the magnetic field about a curved wire~\cite{Saffman1993}. Harmonic fields in the tubular neighbourhood of a knot have also recently been studied in Ref.~\cite{Duan2018}. 

    \subsection{Longitudinal structure --- the solid angle framing}
    
    \begin{figure}[t]
    \makebox[\textwidth][c]{\includegraphics[clip = true, trim = 0 580 0 0, width= \paperwidth ]{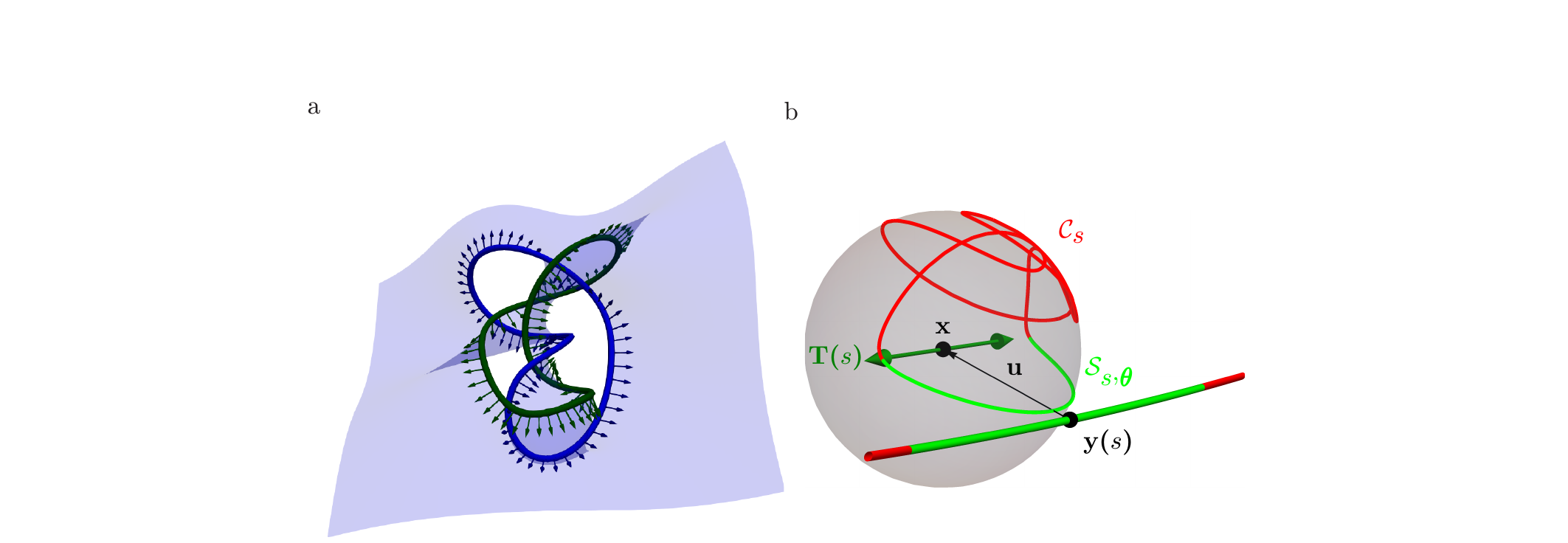}}
    \caption{a) The solid angle framing for the Whitehead link of Fig. \ref{fig:SolidAngle}. Shown is the level set $\omega =0$ (blue surface), and its induced framing (components in blue, green). b) The limiting behaviour of $\bf n$ as $\bf x$ approaches $K$ (shown is the behaviour of $\bf n$ about the marked point on the $4_1$ of Fig. \ref{fig:Knot}). $\bf x$ approaches a fixed point ${\bf y}(s)$ on $K$ such that ${\bf u} := {\bf x} - {\bf y}(s) = (\epsilon \cos\theta, \epsilon \sin\theta,0)$ lies in the normal plane to ${\bf y} (s)$. As $\epsilon/\rho \rightarrow 0$, a region on $K$ of size $\sqrt{ 2 \rho \epsilon}$ (green) projects to a semicircle $\mathcal{S}_{s,\theta}$ between $\pm{\bf T}(s)$. This semicircle sweeps the observation sphere as $\theta$ is varied. The remainder of $K$ projects to $\mathcal{C}_s$ (red), and is independent of $\theta$.} 
    \label{fig:LocalStructure}
    \end{figure}
    
    The intersection of the level set $\omega=0$ with $K$ defines a `solid angle' framing, canonical in the sense that it depends only on the knot and is purely geometric. As this framing is described by a pushoff of $K$ onto an orientable surface, it has zero self-linking number~\cite{Lickorish1997}; the extension to links is straightforward and discussed in \S~\ref{sec:NumericalImplementation}. Fig.~\ref{fig:LocalStructure} a) shows this surface and its induced framing for the Whitehead link of Fig.~\ref{fig:SolidAngle}. A natural question is to identify this solid angle framing in terms of the curve geometry. Let ${\bf x}$ approach a particular point ${\bf y}(s) \in K$, for a fixed $s$, in such a way that the displacement vector ${\bf u} := {\bf x} - {\bf y}(s)$ defines a direction in the normal plane to the curve at $s$ (Fig. \ref{fig:LocalStructure} b)). Aligning the $x,y,z$ axes with the local Frenet-Serret frame ${\bf N}(s),{\bf B}(s),{\bf T}(s)$, we have ${\bf u} = (\epsilon \cos\theta, \epsilon \sin\theta,0)$. As $\epsilon/\rho \rightarrow 0$, where $\rho$ is the radius of curvature, we may think of the image of $K$ on the observation sphere as comprised of two parts; for points ${\bf y}(s')$ with $s'$ outside a small interval $I$ around $s$ (of size $\sim\sqrt{2\rho \epsilon}$), the projection to ${\bf x}$ is no different from the projection to ${\bf y}(s)$, and the image of $K$ is given by the unit chords $\frac{{\bf y}(s') - {\bf y}(s)}{|{\bf y}(s') - {\bf y}(s)|}$. This is a curve $\mathcal{C}_s$ on the observation sphere with endpoints $\pm {\bf T}(s)$ and is independent of $\theta$. In the same limit, the points ${\bf y}(s^{\prime})$ with $s^{\prime}\in I$ contribute to the image of $K$ on the observation sphere a semicircle $\mathcal{S}_{s,\theta}$ between $\pm{\bf T}(s)$ with midpoint $- \frac{{\bf u}}{|{\bf u}|}$ that depends on $\theta$. ${\bf n}$ is thus decomposed as ${\bf n} = \mathcal{C}_s \cup \mathcal{S}_{s,\theta}$. Varying $\theta$, $\mathcal{C}_s$ remains unchanged, and $\mathcal{S}_{s,\theta}$ wraps the sphere once, giving the asymptotic winding structure $\omega = 2(\theta - \alpha(s))$, where $\alpha(s)$ is the rotation angle of the Frenet-Serret normal ${\bf N}(s)$ into the solid angle framing. $\alpha(s)$ gives the longitudinal structure of $\omega$. It represents the contribution of $\mathcal{C}_s$ to $\omega$, and as such is a global quantity, not computable by a local analysis. 

    Our decomposition of ${\bf n}$ is identical to that of the set of cross chords considered in the context of C\u{a}lug\u{a}reanu's theorem~\cite{Dennis2005,Calugareanu1959}, a consequence of the projection map outside of $I$ degenerating to the chord map as $\epsilon/\rho \rightarrow 0$ to give $\mathcal{C}_s$. The completion of $\mathcal{C}_s$ by $\mathcal{S}_{s,\theta}$ is given, in C\u{a}lug\u{a}reanu's theorem, by a choice of framing vector ${\bf u}$ for $K$~\cite{Dennis2005}. Here it is given, via projection, by the displacement vector ${\bf u}$.

    As discussed by Dennis \& Hannay in Ref.~\cite{Dennis2005}, given some framing ${\bf u}$, $\mathrm{Wr}(K)$ and $\mathrm{Tw}(K, {\bf u})$ are given by the areas swept out on an abstract sphere by $\mathcal{C}_s$ and $\mathcal{S}_{s,\theta}$ respectively, as $s$ varies along $K$. They point out that one may choose a special framing, which they call the `writhe framing', such that the area swept out by $\mathcal{S}_{s,\theta}$ precisely cancels that swept out by $\mathcal{C}_s$, giving zero self-linking number. The discussion above makes clear this framing is exactly the solid angle framing, and the cancellation condition may be naturally read as a variation of $\theta$ such that $\frac{{\bf u}}{|{\bf u}|}$ lies tangent to the level set $\omega = 0$; in terms of the Frenet-Serret frame, $\theta= \alpha(s)$.

    \subsection{Transverse structure --- curvature induced corrections to $\omega$}
    
    In the previous section, we saw that the asymptotic structure of $\omega$ normal to $K$, corresponding to the decomposition ${\bf n} = \mathcal{C}_s \cup \mathcal{S}_{s,\theta}$, is simply $\omega = 2( \theta - \alpha(s))$. At finite $\epsilon/\rho$ we find corrections due to the local curvature of $K$, with the leading contribution being logarithmic in $\epsilon$. For the derivative of $\omega$, the magnetic field, this problem is well studied \cite{Saffman1993,Ricca1994}. However, we wish to demonstrate that existing results may be mapped directly on to corrections in the geometry of $\bf n$ as the decomposition ${\bf n} = \mathcal{C}_s \cup \mathcal{S}_{s,\theta}$ is smoothed at finite $\epsilon/\rho$, insight one does not gain from the magnetostatic picture.

\begin{figure}[t]
    \begin{centering}
    \includegraphics[clip = true, trim = 120 580 120 0, width = \textwidth ]{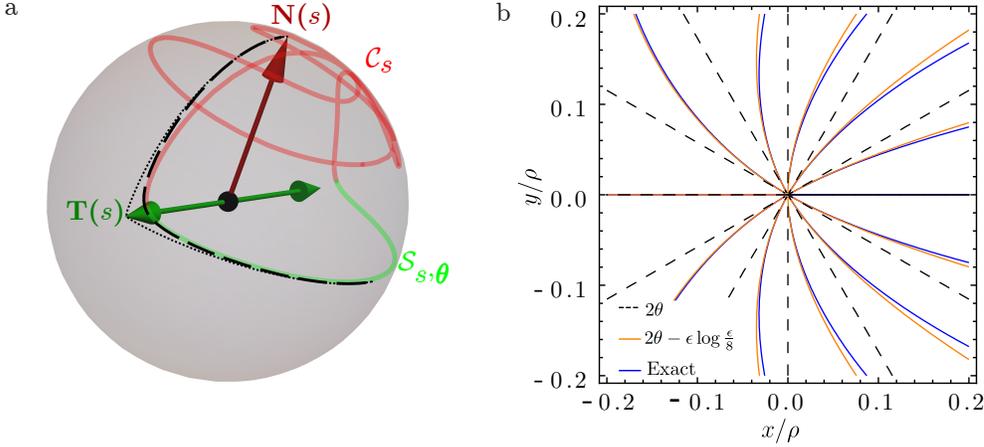}
    \caption{a) For finite $\epsilon/\rho$, the local structure of $\bf n$ is approximated by the hyperbola \eref{eq:CurvatureCorrectedn} --- the dashed black line gives the approximation to $\bf n$ shown in Fig.~\ref{fig:LocalStructure}. As $\epsilon /\rho \rightarrow 0$, the vertex of this hyperbola approaches ${\bf T}(s)$, and the asymptotes (black dotted lines) remain unchanged. In this way, we obtain the limiting decomposition ${\bf n} = \mathcal{C}_s \cup \mathcal{S}_{s,\theta}$. The two asymptotes are great circles through ${\bf T}(s)$ at angles $\theta$ and $0$, and give the local behaviour of $\mathcal{S}_{s,\theta}$ and $\mathcal{C}_s$ --- note that that an angle of $0$ corresponds to the direction ${\bf N}(s)$. b) The local structure of $\omega$ in a plane normal to $K$. Contours of spacing $\frac{\pi}{3}$ are shown for the the zeroth order rotational structure (black dashed line), the curvature induced correction \eref{eq:CurvatureCorrection} (orange) and the exact solution for a circle of radius $\rho$ (blue). The absolute values of the level sets are arbitrary, as we have discarded global information about $\mathcal{C}_s$ in our local structure calculations. The primary effect of curvature is to bunch the level sets of $\omega$ along the local normal. Note that for the curvature induced correction we have fixed the regular values in \eref{eq:CurvatureCorrection} to zero by comparison with the exact solution for a circle \cite{Saffman1993}.}
    \label{fig:CurvatureCorrections}
    \end{centering}
    \end{figure}

    The asymptotic description ${\bf n} = \mathcal{C}_s \cup \mathcal{S}_{s,\theta}$ contains cusps at the boundary between $\mathcal{C}_s$ and $\mathcal{S}_{s,\theta}$, located at $\pm {\bf T}(s)$. The primary effect of small but finite $\epsilon/\rho$ is a rounding of these cusps, and the displacement of ${\bf n}$ slightly off $\pm {\bf T}(s)$, as shown in Fig.~\ref{fig:CurvatureCorrections} a). It is thus natural to focus our attention, and choose coordinates, appropriate to describing $\bf n$ in the vicinity of $\pm {\bf T}(s)$. Expanding ${\bf y}(s')$ to lowest order in $s'$, ${\bf y}(s') = (\frac{1}{2\rho}(s'-s)^2,0,s'-s)$ and $\bf n$ is given by
    \begin{equation}
{\bf n} = \biggl[1 + \frac{\tilde{\epsilon}}{2} \biggr(\tilde{s}^2 +\frac{1}{\tilde{s}^2}\biggl)- \tilde{\epsilon}\cos \theta \biggr]^{-\frac{1}{2}} 
\biggr(
        \sqrt{\frac{\tilde{\epsilon}}{2}}\frac{1}{\tilde{s}} (\tilde{s}^2 - \cos \theta),
        -\sqrt{\frac{\tilde{\epsilon}}{2}}\frac{1}{\tilde{s}} \sin \theta,
        1
        \biggl),
    \label{eq:FirstCurvatureCorrectedn}
    \end{equation}
    where we have defined reduced lengthscales $\tilde{\epsilon}:=\frac{\epsilon}{\rho}$, $\tilde{s} :=\frac{s'-s}{\sqrt{2\epsilon \rho}}$. The form of \eref{eq:FirstCurvatureCorrectedn} is chosen to emphasise that we have an expansion of $\bf n$ in the vicinity of $\pm{\bf T}(s)$ on the observation sphere. Focusing now on the smoothed cusp at positive $\tilde{s}$, we introduce a new variable $t :=\ln(\tilde{s})$, and rotate the $x$-$y$ coordinates of ${\bf n}$ by $\frac{\theta}{2}$, yielding 
    \begin{equation}
{\bf n}
= [1 + \tilde{\epsilon}(\cosh 2t - \cos \theta )]^{-\frac{1}{2}}
\biggr(
        \sqrt{2\tilde{\epsilon}} \cos \frac{\theta}{2} \sinh t, 
        -\sqrt{2\tilde{\epsilon}}\sin \frac{\theta}{2} \cosh t,
        1
        \biggl),
    \label{eq:CurvatureCorrectedn}
    \end{equation}
    a hyperbola projected onto the observation sphere (Fig.~\ref{fig:CurvatureCorrections} a)). In the original, unrotated coordinates, the asymptotic behaviour of this hyperbola is of two longitudinal great circles passing through ${\bf T}(s)$ at angles $\theta$ and $0$. As $\tilde{\epsilon}\rightarrow 0$, the first of these circles gives $\mathcal{S}_{s,\theta}$. The second gives the local structure of $\mathcal{C}_s $, and in particular tells us that the direction of departure of $\mathcal{C}_s$ from ${\bf T}(s)$ is set by ${\bf N}(s)$. The vertex of the hyperbola, found at $t = 0$, is the point of closest approach to ${\bf T}(s)$ and gives the natural choice $\tilde{s} =1$ ($s' = s + \sqrt{2\rho \epsilon}$) to define the upper boundary between $\mathcal{S}_{s,\theta}$ and $\mathcal{C}_s $. It approaches the pole as $\sqrt{{\tilde{\epsilon}}}$, and so in the limit $\tilde{\epsilon} \rightarrow 0$ we recover the sharp decomposition ${\bf n} = \mathcal{C}_s \cup \mathcal{S}_{s,\theta}$.

The local structure of the solid angle can be computed using any of our formulae for $\omega$, however, in view of the foregoing description, an appealing method is to use~\eref{eq:GaussBonnetImmersed} and the geodesic curvature of the hyperbola. As this approach is symmetric in $\tilde{s}$, it is enough to compute the geodesic curvature for the hyperbola near $\tilde{s}=1$ and simply double the result to account for $\tilde{s}=-1$. Further, the geodesic curvature of $\bf n$ is strongly peaked in a localised region of size $\sim\sqrt{\tilde{\epsilon}}$ about the vertex of the hyperbola, decaying to $0$ as the hyperbola approaches its asymptotic great circles. 
Using~\eref{eq:CurvatureCorrectedn} we find an integrated geodesic curvature of 
\begin{equation}
-2 \int_{-\infty}^{\infty}\frac{\sin \theta\sqrt{1+\tilde{\epsilon}( \cosh 2t - \cos \theta)} }{\cos \theta + \cosh 2t + \tilde{\epsilon} \sin^2 \theta } \mathrm{d}t,
\label{eq:CurvatureCorrectionIntegration}
\end{equation} 
where we have extended the upper limit of integration to $+\infty$, corresponding to an integration of the hyperbola between $- \frac{{\bf u}}{|{\bf u}|}$ and ${\bf N}(s)$ on the observation sphere. The integrand decays exponentially for large $t$ so that the error involved is small. 

The integral~\eref{eq:CurvatureCorrectionIntegration} may be evaluated exactly in terms of elliptic integrals of the first and third kind. The main feature is that the result is not analytic in $\tilde{\epsilon}$ but has leading behaviour $\tilde{\epsilon} \ln \tilde{\epsilon}$. This can be seen most easily by noting that the integrand decays exponentially for $|t| \gtrsim \frac{1}{2} \ln (2/\tilde{\epsilon})$ and that the integral is dominated by values of $|t|$ smaller than this. 
Retaining only the leading behaviour, one finds the local structure of the solid angle has the form 
\begin{equation}
\omega(\tilde{\epsilon}, \theta)= 2 \bigl( \theta-\alpha(s) \bigr) + \tilde{\epsilon} \ln \frac{8}{\tilde{\epsilon}}\, \sin \theta + O(\tilde{\epsilon}) ,
\label{eq:CurvatureCorrection}
\end{equation} 
in which a zeroth order term from the integrated geodesic curvature gives the winding of $\omega$ and the logarithmic term causes the level sets of $\omega$ to bunch along the local normal. Fig.~\ref{fig:SolidAngle} d) shows a cross-section through a Whitehead link in which both of these structures are clearly visible. In Fig.~\ref{fig:CurvatureCorrections} b) we compare the various orders of approximation in \eref{eq:CurvatureCorrection} to the exact solution for a round unknot. In contrast to the divergence of the magnetic field, $\omega$ is perfectly well behaved as $\tilde{\epsilon} \rightarrow 0$. The logarithmic correction $\tilde{\epsilon} \log \tilde{\epsilon}$ tends to $0$, but in a cusped manner, with unbounded radial derivative at the origin. We may interpret this fact as a direct consequence of the limiting cusped structure ${\bf n} = \mathcal{C}_s \cup \mathcal{S}_{s,\theta}$ --- the magnetic field gives the rate of change in the area of a spherical curve as we smooth a cusp in it, and is thus naturally unbounded. 

We note briefly that \eref{eq:CurvatureCorrection} is not harmonic --- indeed, the corresponding expression for the magnetic field found in, for example, \cite{Saffman1993} is not divergence free. This is a consequence of neglecting variation in $\omega$ along ${\bf T} (s)$ and one may verify that, allowing $\bf {x}$ to lie off the plane normal to ${\bf y}(s)$, one picks up a term linear in $z$ which restores harmonicity. 

    \section{Remarks on numerical implementation, extension to links}
    \label{sec:NumericalImplementation}

    In \eref{eq:OurSolidAngle1}, \eref{eq:GaussBonnetImmersed}, \eref{eq:DualCurveImmersed} and \eref{eq:OurSolidAngle2}, we have several possible methods for computing $\omega$ for any curve $K$, directly from the specification of its embedding in $\mathbb{R}^3$. The main difficulties in their numerical implementation are encountered when evaluating $\omega(\bf x)$ at points close to the surface of discontinuity discussed in \S\ref{sec:CurveIsotopies}, \ref{subsec:Pullback}. We shall focus discussion on \eref{eq:OurSolidAngle1} and \eref{eq:OurSolidAngle2}, the remaining equations being of similar numerical character.

    Focusing first upon \eref{eq:OurSolidAngle1}, when $\bf x$ lies on the surface of discontinuity it is pierced by a (generically) unique half-line extended from some point ${\bf y}(s) \in K$ such that ${\bf n}(s) \cdot {\bf n}_\infty = -1$. Considering the integral in \eref{eq:OurSolidAngle1} to be defined upon $K$, at the arc length $s$ there is an isolated point of divergence in the integrand. In the degenerate case where ${\bf x}$ lies upon a line of self-intersection in the surface, there will be multiple such points. Letting ${\bf x}$ now lie slightly off the surface and approach it perpendicularly, we may expand the integrand of \eref{eq:OurSolidAngle1} using ${\bf x}-{\bf y}(s) := \epsilon \cos \theta\, {\bf n}_\infty + \epsilon \sin \theta\, {\bf n}_\infty \times {\bf T}(s) / |{\bf n}_\infty \times {\bf T}(s)| $, where $\theta$ is now the angle between $ {\bf x}-{\bf y}(s)$ and the surface. We find that its limiting behaviour is that of a Lorentzian peak of width $\epsilon \theta$, which abruptly switches sign as ${\bf x}$ crosses the surface. If one employs a simple numerical integration scheme with regularly spaced points along $K$ of spacing $\Delta s$, the Lorentzian peak is not captured when $\epsilon \theta \approx \Delta s$. This leads to poor approximation of $\omega({\bf x})$ in a region of constant thickness $\Delta s$ about the surface of discontinuity. By refining $K$, we may reduce the thickness of this region --- unsurprisingly, this result suggests that $\Delta s$ should be on the order of the resolution one desires for $\omega$.
    
    A similar discussion holds for \eref{eq:OurSolidAngle2}, for which the divergences of the integrand occur at $s$ such that ${\bf n}(s)\cdot{\bf T}(s) = \pm 1$, depending on which homotopy is used. The width of the Lorentzian peak instead scales as $\rho(s) \theta$, and so the thickness of the region of poor approximation is $\Delta s \epsilon/ \rho (s)$; in particular, we note that this thickness scales with viewing distance in \eref{eq:OurSolidAngle2}, but not in \eref{eq:OurSolidAngle1}.

    One method of avoiding these peaks is to use the freedom in \eref{eq:OurSolidAngle1}, \eref{eq:OurSolidAngle2} to move the surface of discontinuity about in space, ensuring ${\bf x}$ is never too close to it when computing $\omega({\bf x})$. In \eref{eq:OurSolidAngle1}, we have freedom in our choice of ${\bf n}_\infty$. The surface of discontinuity is given by dragging $K$ to infinity along ${\bf n}_\infty$, and two different choices of ${\bf n}_\infty$ will give two such surfaces. If $K$ is knotted, these surfaces must intersect, giving a set of curves on which a third choice of ${\bf n}_\infty$ is needed. In practice, an initial choice of ${\bf n}_\infty$ is often suggested by the geometry of the input knot, or is simply chosen to be a coordinate axis. When computing $\omega({\bf x})$, one may record the minimum value of ${\bf n}\cdot {\bf n}_\infty$ and, if it crosses some user defined threshold, switch to using $-{\bf n}_\infty$ for the calculation at that point. On the set of lines where this second choice again crosses the threshold, a random choice of ${\bf n}_\infty$ may be used. \eref{eq:OurSolidAngle2} faces analogous problems on the tangent developable surface. Here, we have freedom in whether to place the discontinuity on ${\bf T}_{t,+}$ or ${\bf T}_{t,-}$. However, these two surfaces again generically intersect \cite{Cleave1980,Mond1989}, and there is now no more freedom in $\eref{eq:OurSolidAngle2}$, forcing one to either switch method or analytically correct for the Lorentzian peaks along such intersections. For this reason, and for the scaling properties discussed above, from a numerical standpoint we have found the use of \eref{eq:OurSolidAngle1} to be more convenient than \eref{eq:OurSolidAngle2}. 

    Two brief computational remarks: As discussed in \S\ref{sec:LocalStructure}, the limiting local structure of $\omega$ about $K$ has cylindrical symmetry. If one desires high accuracy to sample the tubular neighbourhood of $K$, one may use a cylindrical mesh out to a distance ${\sim}\rho(s)$. Finally, we note that as values of $\omega$ for different values of $\bf x$ are computed independently of one another, our formulae are easily parallelised. 

    \subsection{Extension to links}
    Extending our results to links is straightforward: by the linearity of electromagnetism, one simply sums $\omega$ mod $4 \pi$ for each component of $K$. We reiterate that $\omega$ is only defined for oriented curves, and that different choices of orientation for each component of $K$ will give distinct solid angle functions. In the case of the solid angle framing discussed in \S\ref{sec:LocalStructure}, each component $K_i$ acquires a framing, whose self-linking number equals the negative of the sum of the linking numbers between $K_i$ and $K_j$, $j \neq i$ (Fig.~\ref{fig:LocalStructure}).

    \section{Construction of knotted fields: two illustrations}

    We describe briefly two different examples of knotted fields that can be constructed using the solid angle as illustrations of how it influences the structure in different settings. 

    \subsection{Scroll waves in excitable media}
    \label{subsec:scroll}

    The possibility of knotting in the waves of excitable media has been considered for some time~\cite{Winfree1983,Winfree1984}. In a three-dimensional excitable medium, scroll waves of excitation emanate from a vortex filament, which it is possible to close into a loop or knot. Recent results have highlighted a remarkable topology-preserving dynamics in these materials~\cite{Maucher2016,Maucher2017,Maucher2018} in which the geometric shape of the vortex filament relaxes and simplifies but without strand crossings, thus preserving the topology. Simple effective curve dynamics seem insufficient to capture the full behaviour, which depends also on interactions mediated by the global structure of the scroll waves. This structure can be captured, in part at least, using the solid angle. 

    \begin{figure}[t]
    \centering
    \includegraphics[width=.65\textwidth]{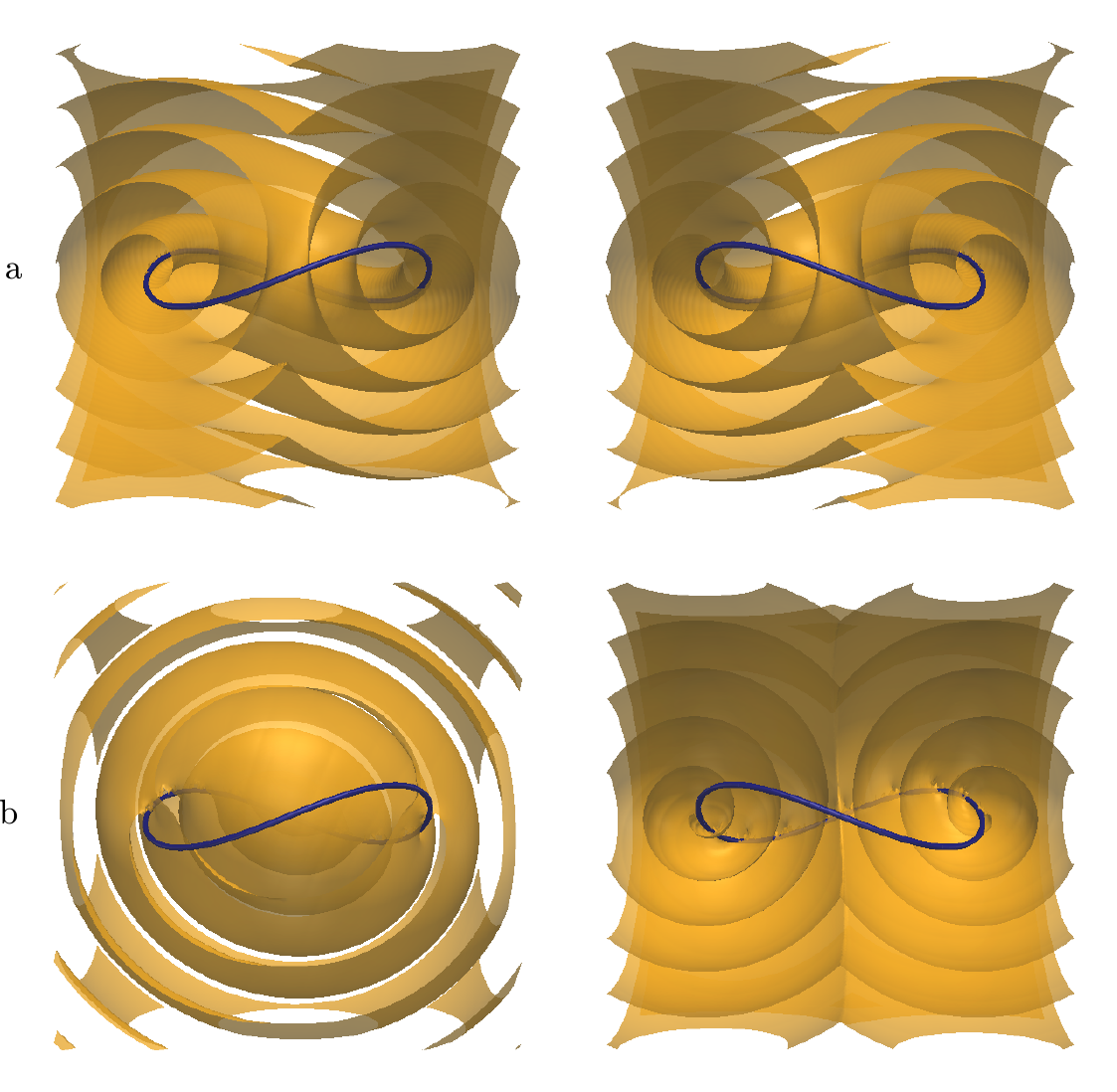}
    \caption{Scroll waves from a unknotted vortex filament. In a) we show the zero level set of the phase field \eref{eq:scroll} and in b) a modification of it where a sinusoidal modulation has been added to the solid angle framing, thereby adjusting the local spin rate of the scroll wave. In both a) and b) the two columns simply show different cuts through the emanating scroll waves.}
    \label{fig:scroll}
    \end{figure}

    Scroll waves emanate from a knotted vortex filament creating an outward propagating family of approximately equi-spaced wavefronts. A simplified description of this wave system is given by a phase field that both winds by $2\pi$ around the filament curve and increases linearly with distance from it. This behaviour is captured by the function 
    \begin{equation}
    \psi({\bf x}) = k d_{K}({\bf x}) + \frac{1}{2} \omega_{K}({\bf x}) \quad \textrm{mod}\; 2\pi ,
    \label{eq:scroll}
    \end{equation} 
    where $\omega_K({\bf x})$ is the solid angle of $K$, $d_{K}({\bf x}) = \min_{{\bf y} \in K} |{\bf y}-{\bf x}|$ is the distance from ${\bf x}$ to the curve $K$ and $k$ is a wavenumber. In Fig.~\ref{fig:scroll} a) we show an example of the scroll waves generated by a simple unknotted vortex ring. 
    Note that the way the wave surface attaches to the filament -- {\sl i.e.} the local spin rate of the scroll wave along the length of the filament -- is determined by the solid angle and, in particular, given by the solid angle framing. Of course, the phase function~\eref{eq:scroll} can be modified to vary this; the modulation can by thought of as a $K$-dependent off-set to the distance function $d_{K}({\bf x})$. An example of such a modulation and how it alters the scroll waves is shown in Fig.~\ref{fig:scroll} b). 

    \subsection{Nematic disclinations}
    \label{subsec:nematics}

    In nematic liquid crystals it is possible to manipulate topological defect lines, called disclinations, so as to create closed loops in the form of any knot or link~\cite{Tkalec2011,Copar2015,Machon2013}. The surrounding liquid crystal texture is an example of a knotted field. 
    The molecular orientation in liquid crystals is described by a unit vector ${\bf d}$ with the nematic symmetry ${\bf d} \sim - {\bf d}$; disclinations are line defects in the director field around which the orientation rotates by $\pi$, or reverses. 
    The solid angle facilitates an explicit construction of a knotted field with this property. For example, the director field 
    \begin{equation}
{\bf d}({\bf x}) = \bigl[ \sin\bigl( \omega_{K}({\bf x}) / 4 \bigr) , 0 , \cos\bigl( \omega_{K}({\bf x}) / 4 \bigr) \bigr], 
\label{eq:planar_nematic}
\end{equation}
encodes $K$ as a disclination line for any choice of knotted curve, or link. This knotted field has two particularly notable properties. First, since the solid angle is harmonic, it corresponds to a critical point of the one elastic constant Frank free energy. Second, the texture is ``planar'', having no $y$-component. 

\begin{figure}[t]
\centering
\includegraphics[width=\textwidth]{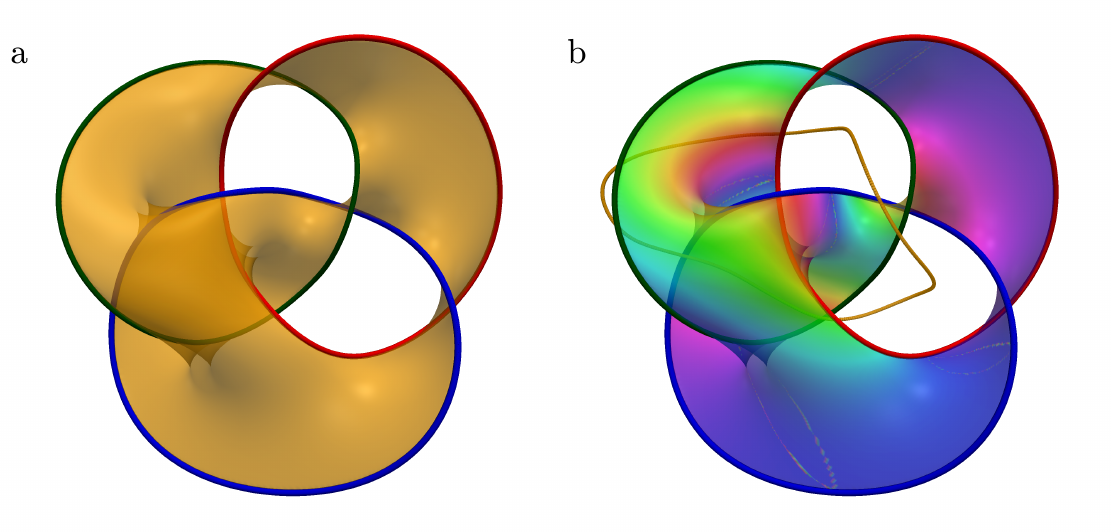}
\caption{Knotted nematic texture for disclinations forming the Borromean rings. The surface corresponds to the set of points where the director has no $z$-component, $d_z=0$; it is coloured according the $xy$-components. In a) the texture is planar (Eq.~\eref{eq:planar_nematic}) and in b) it is fully three-dimensional (the curve defining the $xy$-winding through the angle $\omega_L$ is also indicated).}
\label{fig:nematic}
\end{figure}

We show in Fig.~\ref{fig:nematic} a) a visualisation of the director field \eref{eq:planar_nematic} for the case where the disclination lines $K$ correspond to the Borromean rings. The knotted nematic texture is conveniently visualised by showing the surface where the $z$-component of the director vanishes --- the vector field \eref{eq:planar_nematic} has boundary conditions such that the director is aligned along $z$ asymptotically far from $K$, motivating this choice. This surface is a level set of the solid angle, namely $\omega_{K}=2\pi$. 

A generalisation creating fully three-dimensional knotted nematics is the vector 
\begin{equation}
\fl    {\bf d}({\bf x}) = \biggl[ \sin\biggl(\frac{\omega_{K}({\bf x})}{4}\biggr) \cos\biggl(\frac{\omega_{L}({\bf x})}{2}\biggr) , \sin\biggl(\frac{\omega_{K}({\bf x})}{4}\biggr) \sin\biggl(\frac{\omega_{L}({\bf x})}{2}\biggr) , \cos\biggl(\frac{\omega_{K}({\bf x})}{4}\biggr)  \biggr] ,
    \label{eq:nonplanar_nematic}
    \end{equation}
    where $\omega_{L}({\bf x})$ is a second solid angle function for a curve $L$ chosen as follows. The surface $d_z=0$ is the same as before (the level set $\omega_{K}=2\pi$) but the director field is no longer constant over it, varying with the solid angle function $\omega_{L}$. In Fig.~\ref{fig:nematic} b) we illustrate this through the colour of the surface. Now the gradient of this colour is (proportional to) a magnetic field and $L$ is the curve corresponding to the current carrying wire needed to generate that magnetic field. More formally, $L$ is a curve in the complement of the surface $d_z=0$ corresponding to a homology cycle and generates colour winding around the dual cycle of the surface itself. Knotted nematic fields with any desired topological properties can be constructed in this way but of course the construction is more than purely topological and depends also on the geometric properties of the solid angle and of the curves that generate them. 

    \section{Discussion}
    \label{sec:discussion}

    The solid angle provides a canonical knotted field for any explicitly given curve or link, depending only on that curve and its geometry. As such it facilitates a study of the geometry of knotted fields, shedding light on their structure and establishing connections between the field and the geometry of the curve. 
    We have given a survey of its properties and methods for computing it that parallels and modernises Maxwell's seminal presentation. 
    The fundamental result is the homotopy formula~\eref{eq:Isotopy}, which unifies the different formulae for calculating the solid angle, and also provides the means for characterising changes in the knotted field induced by deformations of the curve. In the latter context, it would be natural to study the consequences of inflection points and other geometric degeneracies in the curve shape, and also strand crossings or, with suitable extension, reconnections. Likewise, one could seek a characterisation of the geometric shape of a knot or link whose solid angle function realises specific properties, for instance having a minimal number of critical points. Those special geometric shapes where the properties of the solid angle change would then represent an interesting branch of singularity theory. 

    The local structure of the field can be considered particularly important in many systems. Here, the natural framing provided by the solid angle and its relation to the writhe of the curve establish a standard reference, from which the global effects of changes to the local behaviour can be systematically assessed.

    \ack{We are grateful to R.B.~Kusner and R.L.~Ricca for discussions about Maxwell's work on solid angle. This work was supported by the UK EPSRC through Grant Nos. EP/L015374/1 (JB) and EP/N007883/1 (GPA). JB acknowledges partial support from a David Crighton Fellowship and thanks R.E.~Goldstein and DAMTP for their hospitality during it.}

    \end{document}